# Energy Spectrum of Primary Knock-on Atoms and Atomic Displacement Calculations in Metallic Alloys Under Neutron Irradiation

Faranak Hatami[1]

[1]University of Massachusetts Lowell, Dept. of Physics and Applied Physics, Lowell, MA 01854, USA, Faranak_Hatami@uml.edu

**Abstract**

**Materials subjected to neutron irradiation experience damage due to displacement cascades triggered by nuclear reactions. This paper presents a practical method to calculate primary atomic recoil events (PKAs), which lead to cascade damage, based on energy and recoiling species. We developed a custom code to identify PKAs and extract their properties using MCNPX and SRIM. This code determines the specifications of recoil atoms from the data provided by the PTRAC card in MCNPX. Consequently, the energy spectrum of PKAs generated through various reaction channels, including elastic/inelastic scattering and transmutations such as (n, α), (n, p), and (n, γ), is calculated. This PKA spectrum is then input into SRIM, which calculates the total number of atomic displacements using the binary collision approximation (BCA) and provides crucial information about the spatial distribution of defects within the irradiated material. Our results indicate that elastic scattering is the predominant reaction, producing most PKAs with energies in the range of several keV. In contrast, inelastic scattering becomes the dominant interaction for generating high-energy PKAs ($\sim E_{PKA}$>1 MeV). Additionally, we observed that the number of Frenkel pairs versus ion energy curves for light particle ion implantation (such as H and He) is significantly smaller than for heavier ions.**

## 1. Introduction

A primary goal for computational simulation of materials in nuclear energy systems, is understanding through modelling of the damage created in materials during neutron and ion irradiation [1], [2]. The amount of damage produced by bombarding particles depends on the projectile type and its energy as well as specifications of target material, which leads to differences in defect clustering, swelling, growth, phase change, segregation, bubble formation and dissolution rate [3]–[5]. During neutron and ion irradiation, some collisions transfer energy higher than the displacement threshold value to the primary knock-on atoms (PKAs). These PKAs are displaced from their original lattice sites which in turn, may cause additional displacements of other atoms to occur.

So, the basic effects of irradiation damage in solids are the displacement of atoms from their equilibrium positions and the formation of point defects, as well as defect clustering due to a damage build-up. Cascade displacement damage from PKAs is one of the serious problems in estimation of damage evolution in structural materials used in fission and fusion reactors under neutron irradiation [6]–[9]. There are different approaches for prediction and calculation of the defect evolution and to study behavior of radiation damage on different scales, such as the binary collision approximation (BCA), the molecular dynamics (MD) and the kinetic Monte Carlo (KMC) [10], [11]. All of these methods require information about the produced PKAs, in the form of the type, energy spectrum and spatial distribution [11] For this purpose, *Gilbert et al.*, used the group-wise incident-to-recoil energy cross section matrices, based on elaborated nuclear data libraries generated using NJOY-12 code system, as well as SPECTRA-PKA code to calculate the PKA spectrum for every possible nuclear reaction channel on pure targets [11].

In this work we develop a code that couples the Monte Carlo simulation codes MCNPX and SRIM for calculation of radiation displacement damage. For this means, the MCNPX which has an extended capability for solving neutron, gamma and electron problems and also the standard cross section libraries are included, is used for calculation of PKAs spectra [12]. Ion deposition profiles in materials exposed to energetic beams of ions is calculated using SRIM [13]–[16]. The calculation of the PKA spectrum is performed for different materials including Zr, Ni, V, Fe and Ni50 at.%Cr compound under neutron-irradiation flux-spectrum predicted for High Flux Isotopes Reactor (HFIR). Note that we also calculate contribution of scattering and absorption reactions such as (n, p), (n, γ), (n, α) in PKA production for the investigated materials.

## 2. Method
### 2.1 Simulating of neutron interactions

The neutron interaction with target nuclei is simulated using MCNPX (Monte Carlo N-Particle code) with ENDF/B-VII.1 cross-section library. MCNPX can track particles from their source of production until they are either absorbed by target nuclei or escape from the simulation box. Depending on their energy, neutrons can interact with target nuclei through various reactions channels such as elastic and inelastic scattering, (n, α), (n, p), (n, γ), etc., in which a fraction of neutrons' energy is transferred to target nuclei which might create recoiled atoms (~ ions). We have used a spherical geometry in MCNPX simulation with a radius equal to $3\lambda$, where $\lambda$ is the maximum mean free path (mfp) of neutrons considered as shown in Fig.1. The isotropic neutron source is also placed at the center of this sphere. Using "PTRAC" module in MCNPX, one can individually track every neutron and extract features of the recoiled atoms in their trajectory such as "recoil energy" and "recoil angle".

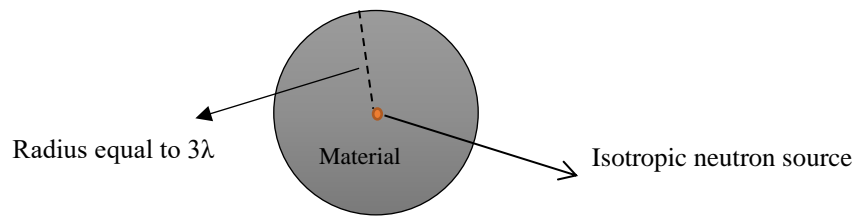

Fig. 1 The geometry of MCNPX simulation with a radius equal to $3\lambda$, where $\lambda$ is the maximum mean free path (mfp) of neutrons considered

PTRAC generates a separate file that includes events, particle types, energy, etc. In this case, details of the recoil ions such as the source (src), bank (bnk), collision (col), capture (cap), and termination (ter) are recorded. These parameters are then employed in later simulations using SRIM. The PTRAC module is completely described in Appendix I of MCNPX official manual [12]. We also developed a code to extract and determine the specifications of PKAs through analyzing PTRAC output. That this code determines the specifications of neutron-induced PKAs and automatically generates an input script named TRIM.DAT which contains the kinetic information about atoms for SRIM calculations. Our home-written code reads the details such as particles' type, energy, position and direction cosines before and after each collision, as well as types of reactions that take place in each event. Table.1 provides a detailed description of different type of events, e.g., we use the event=4000 for calculation of elastic and inelastic collisions, event=2033 for secondary particles from inelastic nuclear interactions. MCNPX contains list of ENDF/B MT numbers for all type of neutron reactions. The MT numbers, discrete reactions, and microscopic cross-section description is indicated in Table 2 [12]

## 2.2 Simulating of recoil ions interaction in SRIM

The interactions of recoiled atoms with target nuclei are simulated using SRIM, which is a software package to study the Stopping and Range of Ions in Matter. SRIM is a Monte-Carlo Binary Collision Approximation (BCA) code for simulating radiation damage, ion collision cascade problems and ion deposition profiles in materials. SRIM can be used to study damage in either one the following modes; (1) ''Ion Distribution and Quick Calculation of Damage,'' and (2) ''Detailed Calculation with full Damage Cascades''. In the Quick Calculation method, the statistical estimates are made based on Kinchin-Pease formalism [10], [14]–[16]; following is the correlation between the number of displacements $v(T)$ and the kinetic energy of recoiled atoms,

$$v(T) = \begin{cases} 0 & E < E_d \\ 1 & E_d < E < 2E_d \\ \frac{T}{2E_d} & 2E_d < E < E_c \\ \frac{E_c}{2E_d} & E \geq E_c \end{cases} \quad (1)$$

Where $E_d$ is the atomic displacement threshold energy and $E_c$ is upper cut-off energy. This mode should be used if the details of target damage or sputtering events are not important. Final distribution of ions in the target, ionization energy loss by the ion into the target, energy transferred to recoil atoms, backscattered ions and transmitted ions can also be calculated in this mode. On the other hand, in "full Damage Cascades" mode, every one of the recoiled atoms is tracked in SRIM until their energy becomes lower than $E_d$. Hence, all the collisional damage to the target is analyzed. Our home-written code determines the specifications of neutron-induced PKAs (~ such as energy, position and recoil angle). Hence, the TRIM.DAT contains atomic number of ions (~ PKAs), their initial energy and starting position and angle, which is then used by TRIM module in "Quick Calculation" mode to calculate the number of total atomic displacements. Fig.2 shows the steps required to find PKA spectra and calculate displacements using MCNPX and SRIM, respectively.

Table 1 Event Type Description of different type of events

| Event Type | | | | | |
|---|---|---|---|---|---|
| src | bnk | sur | col | ter | Flag |
| 1000 | ±(2000+L) | 3000 | 4000 | 5000 | 9000 |

Table 2 MT numbers that describe different types of neutron interactions

| MT number | Reaction type | microscopic cross-section description |
|---|---|---|
| 2 | Elastic | Elastic scattering cross section for incident particles |
| 51, ... , 90 | Inelastic | Production of a neutron, leaving the residual nucleus in the excited state |
| 102 | (n, γ) | Production of an gamma particle. |
| 103 | (n, p) | Production of an proton particle. |
| 104 | (n, d) | Production of an deuteron particle. |
| 107 | (n, α) | Production of an alpha particle. |

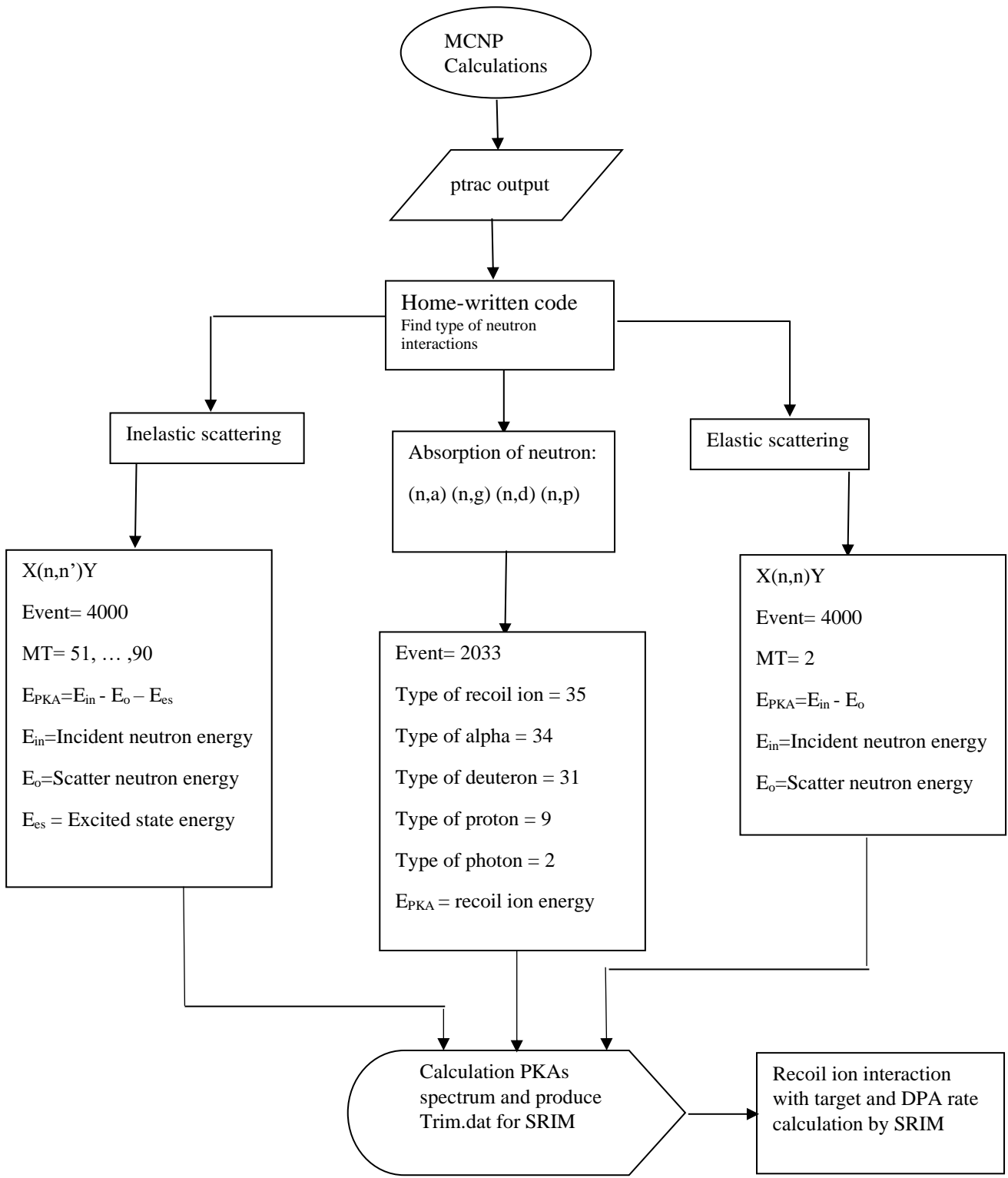

Fig. 2 The steps required to find PKA spectra and calculate displacements using MCNPX and SRIM are shown.

## 3. RESULTS AND DISCUSSION

The PKA spectra for pure Zr, Ni, V and Fe and Ni50at.%Cr compound under neutron-irradiation flux-spectrum predicted for High Flux Isotopes Reactor (HFIR) have been calculated using MCNPX and is shown in Fig. 4. The HFIR spectrum is shown in Fig. 3 [17]. Note that the results are normalized over neutrons emitted from the source (~ nps in MCNPX). To have a better understanding of mechanisms involved in PKA production, following the calculated PKA spectra is divided into three distinct regions and discussed in details:

- Low-energy PKAs (~ approximately $E_{PKA}$<1 keV):

Considering first region, it is obvious that the elastic scattering is the dominant reaction and most of the PKAs are produced with energies as much as a few hundreds of eV with approximately a flat behavior on logarithmic scale, as shown in Fig. 4. This is in agreement with previous report of *Gilbert et al.* [11], in which they also observed an absolute dominance of elastic scattering over other reactions in low energy ranges. In addition, our results show that the inelastic scattering also weakly contributes in PKA production. For example, the average magnitude of PKAs created by elastic scattering in this energy region is approximately 0.45 and 0.36 $^{\#}/_{s.p}$ in pure Fe and Zr, respectively; where this reduces to 0.0024 and 0.000065 $^{\#}/_{s.p}$ considering inelastic scattering. Note that we also observe no contribution of absorption reactions such as (n, p), (n, γ), (n, α) in PKA production in this energy region for the investigated materials. So, it is clear that most of the PKAs are created through elastic and inelastic scatterings and hence, have similar properties to base materials. In the next section, we will further investigate the total damage production due to self-ion irradiation using SRIM.

- Moderate-energy PKAs (~ approximately 1 keV <$E_{PKA}$<1 MeV):

In this region with increasing PKA energy, the elastic scattering contribution exponentially decreases and the inelastic scattering slowly becomes dominant, as shown in Fig. 4. For example, the average magnitude of PKAs produced by elastic and inelastic scatterings are $3.4\times10^{-4}$ and $7.5\times10^{-4}$ $^{\#}/_{s.p}$ in pure Zr, respectively, showing a remarkable increase in contribution of inelastic scattering when being compared to $E_{PKA}$<1 keV region. Moreover, we observe that (n, p) and (n, γ) reactions also contribute in PKA production in this energy region, but the number of PKAs produced through these channels are significantly lower than that of the elastic and inelastic scatterings, e.g., the average magnitudes of PKAs created by (n, p) and (n, γ) are $4.3\times10^{-6}$ and $3.3\times10^{-7}$ $^{\#}/_{s.p}$ in Ni which is much lower than the elastic (inelastic) scattering contribution (~the average magnitudes of PKAs produced from elastic and inelastic are $1.3\times10^{-5}$ and $2.4\times10^{-5}$ respectively). This is also qualitatively consistent with *Gilbert et al.* [11], which showed that the absorption interactions have insignificant effect on the total PKA spectra due to their relatively lower cross-sections in comparison to the elastic scattering. Furthermore, although the results imply that the recoiled atoms through (n, p) reactions are not important in forming the total PKA spectra, but since the hydrogen production and accumulation (~ which eventually leads to formation of bubbles) could have deleterious effects on thermo-mechanical properties of metals on a longer timescale, we have considered the gas-production interactions in our present calculation. Finally, in this energy region, we observe no PKAs produced through (n, α) interaction.

- High-energy PKAs (~ approximately $E_{PKA}$>1 MeV)

Considering Fig. 4, it is clear that the inelastic scattering becomes the dominant interaction in producing high energy PKAs. We also observe no contribution from elastic scattering and absorption interactions including (n, p) and (n, γ) in PKA production in this energy region.

The average magnitude of PKAs created by inelastic scattering is also lower than that of other discussed regions. For example, the average magnitude of PKAs produced by inelastic scattering is 0.36, and $7.5\times10^{-4}$ $^{\#}/_{s.p}$ in first and second region in pure Zr, respectively, where this reduces to $1.2\times10^{-6}$ $^{\#}/_{s.p}$ considering third region (~ high-energy PKAs).

Moreover, in this energy region, we observe that (n, α) interaction slightly contributes in forming the PKA spectra and leads to production of highly energetic recoiled atoms with an average kinetic energy of 7 MeV. This reflects the fact that the PKAs created by this interaction receive a rather high proportion from the neutron spectrum (~ see Fig. 3). For instance, the average magnitude of PKAs created by (n, α) interaction is $6.4\times10^{-7}$ and $1.3\times10^{-7}$ $^{\#}/_{s.p}$ in pure Ni and V with a distribution culminating around 7 MeV, as shown in Fig. 4(c) and Fig. 4(b). These highly energetic recoils can, in turn, severely damage the irradiated structure on the primary damage scale and induce swelling due to accumulation of the insoluble helium and formation of nano-metric bubbles on a long-term damage perspective.

Fig. 4 (e) also shows the PKA spectra for Ni50at.%Cr compound under HFIR neutron irradiation. It is clear that in low-energy region the PKAs are primarily created by elastic scattering interaction, in agreement with our results for pure Zr, Fe, V and Ni (the analysis and calculations presented above). For example, considering $E_{PKA}$ <1 keV, the average magnitude of PKAs created from elastic scattering is $1.3\times10^{-1}$ and $5.6\times10^{-1}$ $^{\#}/_{s.p}$ for Cr and Ni, respectively. Furthermore, our results indicate that the PKAs produced through this channel in Cr are slightly lower than that of Ni implying an approximately identical neutron elastic scattering cross-section. Note that in Fig. 4, the recoil events with kinetic energies less than 1 eV are also neglected. These atoms cannot overcome the atomic binding forces and so, have no contribution in total damage production. In middle-energy region for Ni50at.%Cr compound, inelastic scattering slowly becomes dominant and elastic scattering exponentially decreases, where the PKAs from inelastic scattering for Cr are higher than Ni, e.g., average magnitude of PKA from inelastic scattering is $4.3\times10^{-5}$ and $3.6\times10^{-6}$ $^{\#}/_{s.p}$ for Cr and Ni. However, in high-energy region the PKA from inelastic scattering in Ni are dominant and the PKA through (n, α) interaction from both of the Ni and Cr has negligible contribution in the total PKA spectra.

Moreover, as a representative example, Fig. 5 shows the total PKA spectra as a function of PKA energy in Zr and Fe under HFIR neutron irradiation. The total PKA spectra involves contributions from both elastic and inelastic scatterings and the absorption channels such as (n, γ), (n, α), (n, p). It is obvious that the PKA spectra obtained from the Fig. 4 and Fig. 5 are generally similar behavior. Note that in low-energy range (~ $E_{PKA}$ <10 keV) there is no significant difference between PKA spectrums of Zr and Fe. We also observe that the PKA produced in pure Fe is slightly higher than Zr in high-energy region (~ $E_{PKA}$ >10 keV), where for example, the total average magnitude of PKAs produced in this energy range is $4.55\times10^{-5}$ and $2.36\times10^{-5}$ $^{\#}/_{s.p}$ in pure Fe and Zr, respectively. Considering our present results (~ Fig. 4 and Fig. 5), it can be concluded that the elastic scattering is the dominant reaction and most of the PKAs are produced with energies as much as several hundreds of eV. But, in this paper, to have a more accurate approximation of damage in materials, we consider contribution of both neutron scatterings and absorptions in PKA production.

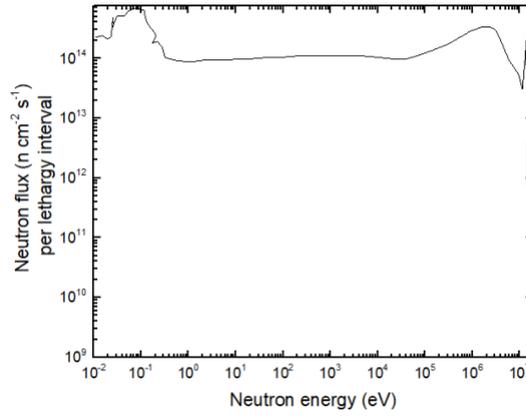

Fig. 3. Neutron irradiation spectra for the HFIR experimental fission reactor [17]

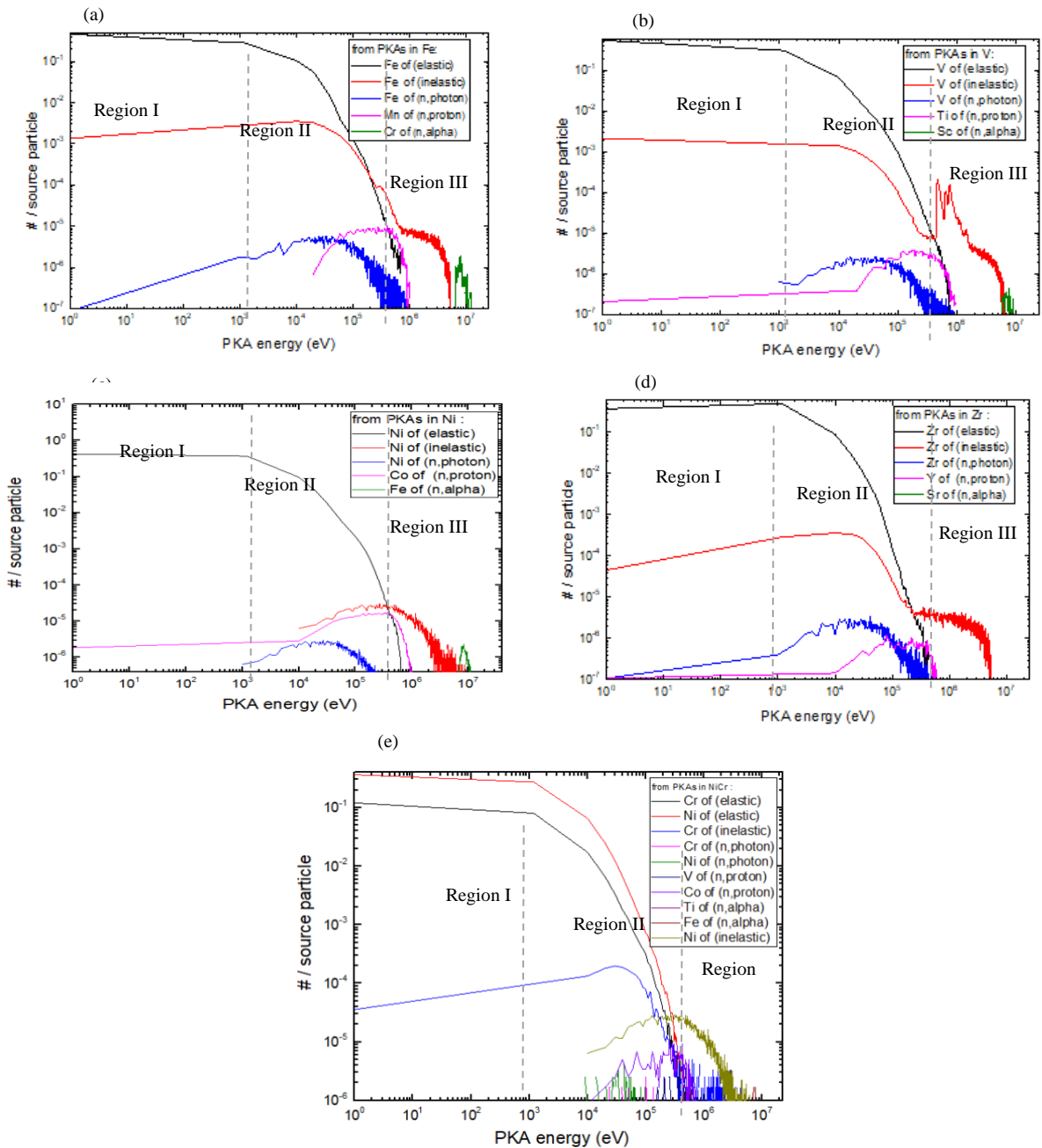

Fig. 4. The PKA spectra for the dominant reaction channels in (a) Fe, (b) V, (c) Ni, (d) Zr and (e) NiCr, under HFIR conditions. (n,photon) = (n, γ) , (n,alpha) = (n, α) , (n,proton) = (n, p)

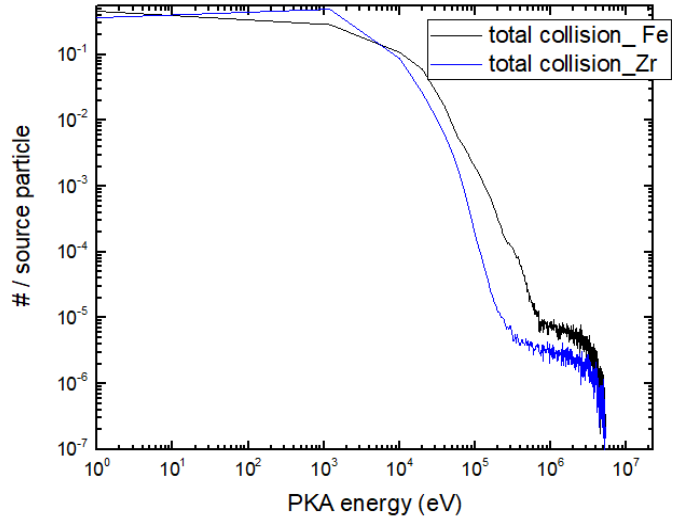

Fig. 5. The PKA spectra for the total of reaction channels in Fe and Zr under HFIR conditions.

The distribution of total number of PKAs, as a function of distance from the center of target, is shown in Fig. 6, which illustrates the distance from center of target increasing, the PKAs formation ratio decreases. The results imply that the total number of PKAs are 828 and 108 at the distance of $4.5 \times 10^9$ Å for Zr and Fe, respectively. Note that for both Fe and Zr, the distribution of PKAs illustrates the more important for the closer distances from the center of target.

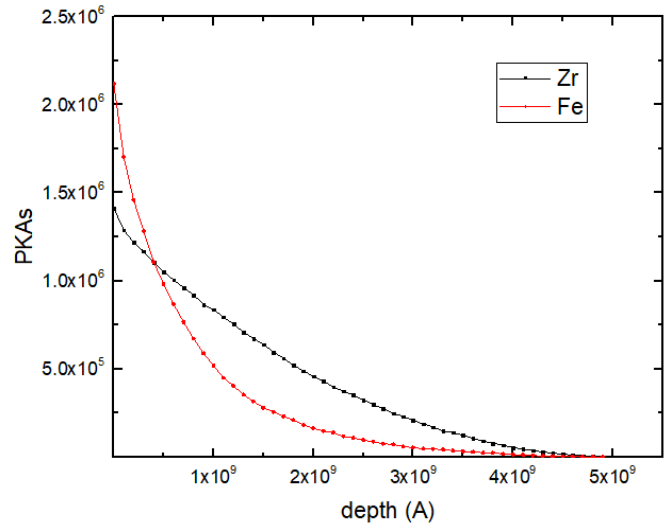

Fig. 6. Production rate of PKA vs depth (Å) in Fe and Zr

As mentioned above, the PKA spectra from the neutron interaction with target is simulated using MCNPX with ENDF/B-VII.1 cross-section library. But M.R. *Gilbert et al*, used the TENDL-2014 cross section library in NJOY-12 code system. Note that for the comparison and validation our simulation, we have used the TENDL-2017 cross section library in MCNPX simulations.

The energy spectrum of PKAs produced through various reactions channels such as elastic and inelastic scattering in pure Zr and Fe under neutron-irradiation flux-spectrum predicted for HFIR have been calculated using MCNPX with TENDL-2017 cross section library and the results also obtained from the SPECTRA-PKA code, where developed by *Gilbert et al,* is shown in Fig. 7. It illustrates that in low-energy region (~ approximately $E_{PKA}$<10 keV): the PKAs are primarily created by elastic scattering interaction, in agreement with previous section. Also in high-energy PKAs (~ $E_{PKA}$>0.1 MeV), the inelastic scattering becomes the dominant interaction. However, there is some little difference between simulations that's because have been used of two types simulation codes (NJOY-12 and MCNPX). Our simulations with TENDL-2017 cross section are in agreement with the PKA spectra obtained from the SPECTRA-PKA code. The figure shows that there is a valid approximation for our work.

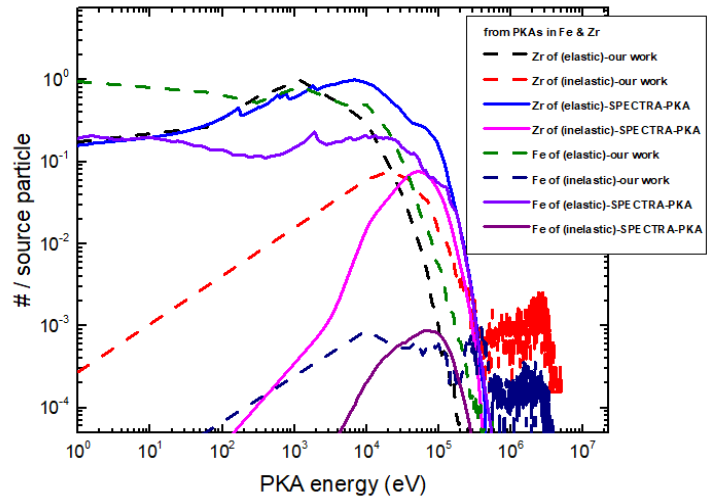

Fig. 7. The PKA spectra for the dominant reaction channels (elastic and inelastic) in Zr and Fe under HFIR conditions is simulated using MCNPX with TENDL-2017 cross-section library and *Gilbert et al* used the TENDL-2014 cross section library in NJOY-12 code. The dashed and full lines illustrate our simulations and *Gilbert* results, respectively.

- Damage calculations in SRIM:

In this section, we present the Frenkel pairs (FPs) production rate as a function of ion (~ PKA) energy using SRIM. SRIM can be calculated the total number of displacements per atom (DPA) resulting from PKAs. TRIM.DAT script –obtained by MCNP and home-written code– is used as input in the SRIM and radiation damage calculations in the Quick Calculation mode. Authors are recommending that run SRIM using Quick Calculation option to compute displacements per atom (see Ref. [1] for details) as well as E. Lagzdina et al. [18] used both Quick Calculation and Full-Cascades modes of SRIM code. The total numbers of obtained displacements don't much different using both of the modes. As mentioned earlier, the TRIM.DAT includes particles type, energy, position and direction cosines. Then, this input script is used by SRIM in Quick Calculation mode for calculating neutron induced displacement damage parameter. The total number of target displacements (Frenkel Pairs) caused by PKAs interaction with different materials such as Zr, Fe, Ni and V as target simulated by SRIM are summarized in Table 3.

Considering the results of Table 3, it is obvious the total target displacements computed for the main reaction channels on Fe, Zr, V and Ni have a same qualitative behavior. As shown in Table.3, the number of total target displacements produced by PKAs of the original host elements rather more than other elements. For example, the total number of displacement created by Zr and Sr are $1.003\times10^{11}$ and $3.8\times10^5$ into Zr. Also illustrates the damage produced by the secondary particles from elastic and inelastic scattering greater in comparison with other interactions such as (n, α), (n, p), that this is in agreement with previous section. As discussed in Fig.4 the light particles are emitted with a higher proportion of the recoil energy by (n, α) interaction (~ approximately about a few MeV), but their total magnitudes lower than elastic scattering. On the other hand, Table.3, indicates that the total displacement caused by Fe in Ni is $3.2\times10^7$, that this value is approximately about a few hundred times lower than the value from Ni into Ni. Note also, the total displacement created in V by (n, p) interaction is $8.5\times10^6$ and have a little more magnitude than the (n, a) interaction.

We also calculated the Number of Frenkel Pairs ($N_{FPs}$) as a function of PKA (~ion) energy predicted using SRIM code. Fig. 8 shows the $N_{FPs}$ distributions obtained by ion implantation types for simulation of PKAs in Zr and V. Indeed, the number of Frenkel Pairs have been calculated for the range of ion energy and each PKA type predicted under HFIR neutron irradiation, for example, H, He, Zr, Sr and Y into Zr. The resulting $N_{FPs}$ versus ion energy curves for ion implantation such as light particles (~H and He) into Zr and V have a significantly smaller size than the other heavy ions implantation. The results suggest that the rate of Frenkel pair's production is rather an order of magnitude higher in host elements. For example, in Fig. 8(e) and Fig. 8(f), the average of NFPs created by H and Zr implantation in Zr are of the order of 10 and $10^4$, respectively. Therefore, there is a significant difference between the NFPs produced of the H and Zr implantations, as shown in Fig.8.

Table. 3. Damage calculations for MCNP output using SRIM

| Incident ion (PKA) | target atom | PKA from of | Total Target Displacements (for 10000 atoms) |
|---|---|---|---|
| Zr | Zr | (n, n) | $1.003\times10^{11}$ |
| Sr | Zr | (n, α) | $3.8\times10^5$ |
| Y | Zr | (n, p) | $2.5\times10^6$ |
| Ni | Ni | (n, n) | $3.86\times10^9$ |
| Fe | Ni | (n, α) | $3.2\times10^7$ |
| Co | Ni | (n, p) | $1.15\times10^8$ |
| V | V | (n, n) | $2.8\times10^{10}$ |
| Sc | V | (n, α) | $8.5\times10^5$ |
| Ti | V | (n, p) | $8.5\times10^6$ |
| Fe | Fe | (n, n) | $1.630\times10^{10}$ |
| Cr | Fe | (n, α) | $4.08\times10^6$ |
| Mn | Fe | (n, p) | $1.9\times10^7$ |

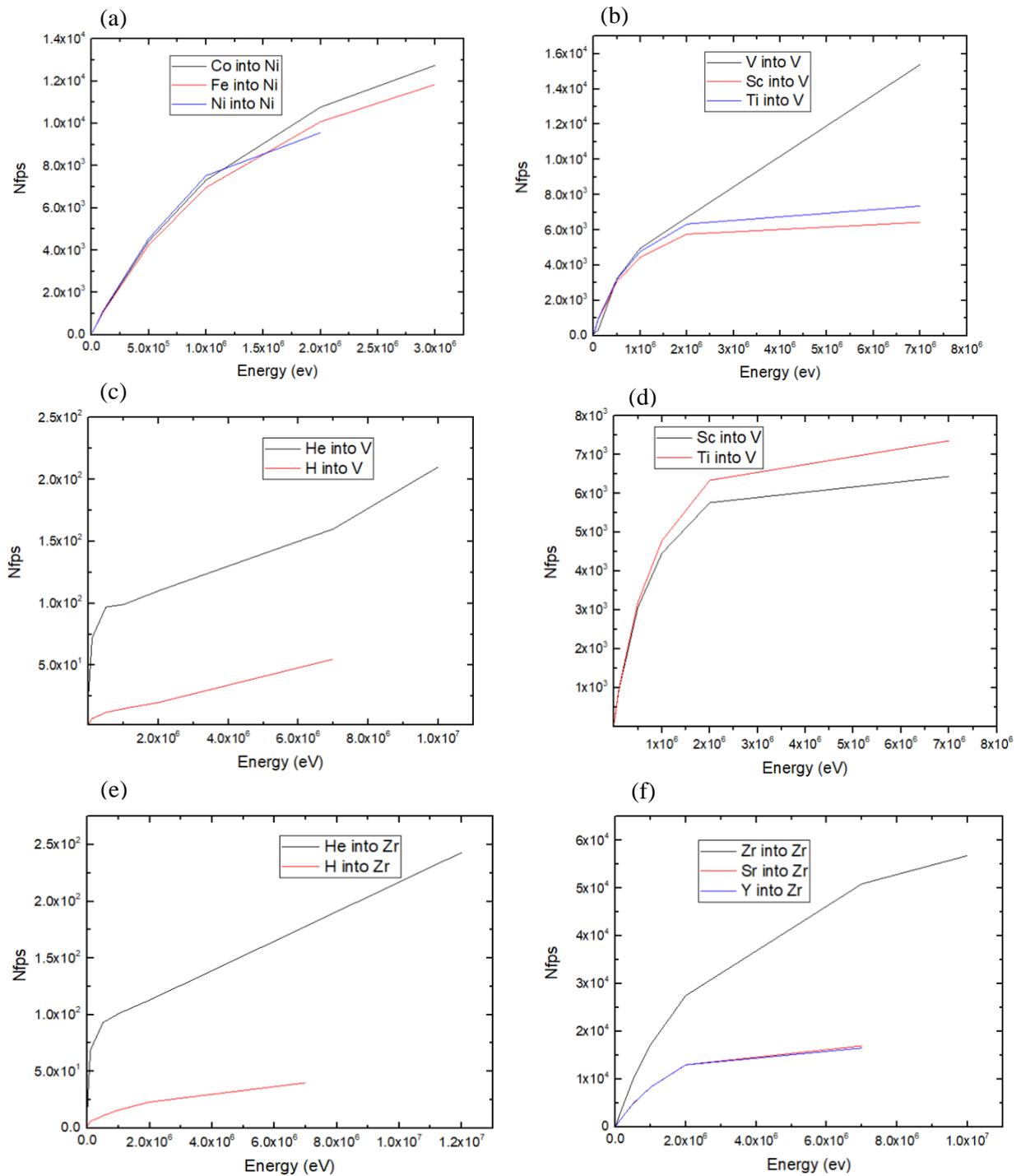

Fig. 8. Number of Frenkel Pairs ($N_{FPs}$) distributions as a function of ion energy predicted using the BCA code SRIM. The $N_{FPs}$ have been calculated for the range of ion energy and each PKA type predicted under HFIR neutron irradiation, for example H, He, Zr, Sr and Y into Zr.

## 4. Conclusion

The investigation of the primary knock-on events generated during neutron irradiation is a vital component in the understanding of radiation damage. The goal of this paper is to predict the energy spectrum of primary knock on atoms (PKAs) produced through various reactions channels such as elastic and inelastic scattering as well as (n, α), (n, p) and (n, γ) collisions. A home-written program is developed to extract specifications of PKAs using PTRAC module in MCNPX. Our home-written program determines the details of neutron-induced PKAs and automatically generates an input script for SRIM, which is called TRIM.DAT. Next, SRIM uses the TRIM.DAT script as input which contains the kinetic information of atoms. Considering our results, for the materials investigated in this work, it can be concluded that the elastic scattering is the dominant reaction and most of the PKAs are produced with energies as much as several of KeV through this channel. However, in this paper, to find a more accurate approximation of damage in materials, we consider the contribution of both neutron scatterings and absorptions in PKA production. Additionally, displacements are predicted to be generated by PKAs of the original host elements rather than other elements.

The PKA spectra for the total of reaction channels in Fe and Zr under HFIR conditions are significant below hundreds of keV energy range. Using the presented methods in this work, can be predicted the role of the different interaction channels in the produced DPA. Based on this, can be obtained the accurate conception of the radiation damage.

In summary, if a researcher intends to use MCNPX and SRIM to compute PKAs spectra and DPA from an irradiation experiment, the following steps should be complied with:

(1) Choose the displacement threshold energy for different materials including impurities
(2) Run MCNPX using PTRAC module,
(3) Extract the information about PKAs from PTRAC file and build the input file for TRIM code,
(4) Run TRIM module in "Quick Calculation" mode to calculate number of total atomic displacements